# Ozonation of Dielectric Fosters Self-Healing Efficiency in Metalized-Film Capacitors: Quantum-Chemical Simulation

Nadezhda A. Andreeva[1], Cuixia Liu[1] and Vitaly V. Chaban[2]

(1) Peter the Great Polytechnic University, Saint Petersburg, Russian Federation. E-mail: andreeva_na@spbstu.ru.
(2) Independent Scientist, Saint Petersburg, Russian Federation.

**Abstract:**

Metalized-film capacitors (MFCs) employ polymer organic dielectrics like polypropylene (PP) and polyimide (PI), in which self-healing is seen as a key advantage. However, the performance of self-healing depends on specific chemical mechanisms involved. The formation of semiconductive carbonaceous soot represents a critical failure risk. This study investigates how oxygen atom impregnation through ozonation of the dielectric material tunes the composition and electrical conductivity of breakdown products in the PP and PI systems with aluminum-zinc electrodes. We revealed, at the atomistic level, that oxygen atoms tend to remove a fraction of carbon atoms from the semiconductive soot by oxidizing carbon into carbon monoxide in both polymers. In PP, oxygen fraction linearly increases gas mass fraction, thereby reducing soot fraction. In PI, the gas/soot ratio effect of oxygen content is less drastic, still clearly positive. The PP soot conductivity decreases uniformly as larger fractions of oxygen atoms are added. In turn, the PI conductivity drops to ~1500 S/m quickly. The PI soot exhibits narrower band gaps compared to that of PP. The oxygen fraction non-monotonically tailors band gaps, which generally increase. To summarize, ozonation enhances MFC reliability by increasing gas species fraction and reducing soot conductivity. We hereby provide numerical molecular-level insights to rationalize self-healing performance enhancement through polymer ozonation.

Keywords: metalized film capacitor; ozonation; self-healing, polypropylene; polyimide.

**Research Highlights**

Excess oxygen increases gas-phase mass fraction after dielectric breakdown.

The soot conductivity is suppressed as the oxygen content rises.

The PI soot exhibits slightly higher conductivity than the PP soot.

**Introduction**

Metalized film capacitors (MFCs) represent passive electronic components characterized by their use of thin polymer films as dielectrics, which are metalized with vapor-deposited electrodes typically made of aluminum, zinc, or more often their alloys.[1-2] These capacitors are renowned for their compact dimensions, excellent frequency characteristics, and self-healing properties, making them indispensable across a vast array of modern applications. Their unique combination of properties is essential in power electronics — such as in DC-link circuits, snubbers, and motor drives — as well as in renewable energy sources. Furthermore, they are increasingly vital in the advancing field of automotive electronics and in high-precision instrumentation where reliability and performance are paramount.[3]

A defining feature of MFCs is their ability to autonomously recover from localized dielectric breakdowns through an intrinsic process called self-healing.[4-6] During an overvoltage event, a short-circuit at a microscopic defect site causes the metalized electrode (≈0.02–0.1 μm) to vaporize instantaneously due to intense, localized arc-induced heating. This effectively isolates the electrical fault and restores functionality without leading to catastrophic failure. It is important to note that this process involves the evaporation of not only the metal electrode but also a small amount of the surrounding polymer dielectric, which produces gaseous by-products like hydrogen, carbon monoxide, and rarely carbon dioxide. The efficient dissipation of these gases prevents the formation of conductive carbon deposits that could otherwise compromise the long-term insulation resistance and overall capacitor integrity.[7-8]

Ozonation of polymers is used to introduce oxygen-containing functional groups, such as carbonyls and hydroxyls, into insulating materials. This oxidation increases surface polarity and can modify dielectric behavior by raising the dielectric constant and improving interfacial interactions. When applied in a controlled manner, ozonation enables chemical activation of polymer surfaces, allowing tuning of electrical and physicochemical properties without altering the material's basic composition.[9-10]

Among the various dielectric materials available, biaxially oriented polypropylene (PP) is overwhelmingly favored in high-performance MFCs due to its exceptional and well-balanced electrical properties.[11-12] These include an extremely low dissipation factor (tan δ ≈ 0.0002–0.0005), which minimizes energy losses; high intrinsic dielectric strength (up to 650 V/μm), and promising stability over time as compared to the existing competitors. Despite exhibiting these baseline dielectric properties, the relentless push for higher power density, miniaturization, and operation in harsher environments made PP the subject of extensive modification research. The development of nanomaterials based on PP seeks to enhance its exploitative characteristics, such as its dielectric strength, thermal stability, and energy density.[13-14]

The breakdown strength of PP can be substantially enhanced by incorporating nanoparticles that act as charge traps. According to the trap theory, these dopants introduce localized energy states within the insulator band gap, which effectively restrict the movement of charge carriers and mitigate impact ionization. For instance, studies have shown that the addition of 1 wt% nano-$Al_2O_3$ increased the trap depth by 17 % and raised the DC breakdown strength by 37 %.[11, 15] Doping with 0.5 wt% nano-structured ZnO resulted in a 37 % improvement in direct-current breakdown strength and a notable 17 % increase in alternating-current breakdown strength[16]. Advanced chemical modification techniques have also proven highly effective. The irradiation cross-linking of PP film, when modified with pentaerythritol triacrylate $C_{14}H_{18}O_7$, achieved a cross-linking degree of 25.68 %. This cross-linking enhanced the stability of the molecular chains at high temperatures, reduced thermal breakdown frequency, and provided a 72.5 % reduction in conductivity.[17] Research on surface grafting with maleic anhydride $C_4H_2O_3$ combined with bulk cross-linking was shown to significantly enhance the dielectric properties of the PP film. The authors reported a multi-order reduction in conductivity and a 14.8 % increase in breakdown strength.[18]

A possible polymer in MFCs is polyimide (PI), which emerged as a candidate material for overcoming the performance bottlenecks of traditional materials. The high-temperature energy density performance further underscores the great potential of PI dielectric film capacitors for applications under extreme conditions.[19] PI is synthesized via the polycondensation reaction of dianhydrides and diamines. The rings of the main chain endow PI with superior chemical and thermal stability. Aromatic PI exhibits a thermal decomposition temperature around 600 °C. The long-term service temperature reaches 300 °C. An alicyclic PI-based material was recently proposed. This PI maintains a discharged energy density of 4.54 J/cm³ and a charge-discharge efficiency exceeding 90% at 200 °C, outperforming some previous materials.[20] Even after four electrical breakdown cycles, it retains 93% of its initial dielectric breakdown strength. Min et al fabricated epoxy resin-impregnated PI films by simply immersing bare PI films in an epoxy resin solution, followed by curing. At 150 °C, after 5 electrical breakdown cycles, the PI films maintain a competitive breakdown strength and large discharged energy density.[21] Through designing disulfide bond exchange reactions, the insulating film containing two diamine monomers retains a certain degree of self-healing properties even after being subjected to mechanical/electrical damage.[22] $C_4H_{12}N_2S_2$ was incorporated as a self-healing functional monomer into the molecular structure of conventional polyimides. The introduction of $C_4H_{12}N_2S_2$ enabled the film to maintain high transmittance, > 87 %, and tensile strength, > 99 MPa.[23]

Many of the effective polymer modifiers incorporate oxygen within their molecular structure. This observation led our research to specifically investigate the influence of varying [O] content on the products formed during dielectric breakdown and the subsequent cooling phase. The carbonaceous soot exhibits semiconducting properties. This conductive soot can form a permanent bridge between the capacitor's electrodes, ultimately leading to a catastrophic short-circuit failure. A critical factor in mitigating this risk is the composition of the breakdown products. A higher yield of gaseous products inherently results in a smaller volume of residual solid soot, thereby reducing the probability of a conductive path formation. We start from the known chemical compositions of PP and PI and systematically vary the oxygen context to isolate its effect. To accurately simulate the real-world environment of a metalized film capacitor, we included atoms from the Zn/Al electrode (3 Zn atoms and 1 Al atom). Such an addition of metal atoms well mimics common alloyed electrodes in experimental setups.

**Methodology and Methods**

The goal of the modeling is to obtain the most thermodynamically favorable compositions and geometries of the soot after the dielectric breakdown event at 5000…7000 K. At these temperatures, the activation barriers for all possible chemical reactions in PP- and PI-based systems do not limit the reactions. The identities of products are determined by thermochemical stabilities only, i.e., the most stable molecules and fragments are the most frequent products. Hence, to determine plausible products of the dielectric breakdown in every system, the decomposition of each polymer and electrode was carried out 250-400 times, starting from random sets of coordinates and momenta.

The kinetic energy injection method explores the potential energy landscape of the breakdown products by periodically providing an excess kinetic energy via a random distribution of atomic momenta.[24-27] The energy of 5000 K, which mimics the extreme conditions at the breakdown epicenter, was converted to a Maxwell-Boltzmann distribution and added to the existing momenta in the simulated system at room temperature. After the modification of momenta, the PM7-MD simulation was continued with the Berendsen thermostat's target temperature set to 300 K and relaxation time set to 1000 fs. Such a methodology allowed one to relatively quickly remove the excess velocities from the system, while driving its Cartesian coordinates toward the thermodynamically stable chemical structures. The remaining structural perturbations were removed by minimizing the system potential energy by means of the BFGS algorithm. The resulting geometry is a stationary point of the simulated system. Unless any imaginary vibrational frequency was detected, the stationary state is a minimum state at the potential energy surface. The lowest-energy minimum state is a global minimum state. The geometries of the global minimum state and also low-energy local minimum states contain the most probable dielectric breakdown decomposition products.

The high perturbation kinetic energy, corresponding to 5000 K, is used by the studied system to cross any associated kinetic barriers and spontaneously arrive in the state of the most stable breakdown products.[28] The gradual decrease of kinetic energy in the system is useful for products to form the strongest covalent bonds. We did not use 6000 or 7000 K as a perturbation since, according to our previous simulations, no diatomic molecules of our interest emerge at so high levels of entropy.[29] All procedures related to potential energy surface navigation were fulfilled in the GMSEARCH (version 2024) software developed by V.V.C. MOPAC (version 18, https://github.com/openmopac/mopac) was used to compute energies of formation of the investigated systems. VMD (version 1.9.3)[30] and XCrySDen [31] were used for the visualization of molecular structures and trajectories.

The lowest-energy structures of the non-volatile (soot) and volatile (gas molecules) products were additionally optimized using a plane-wave density functional theory. The plane-wave density functional theory (PWDFT) calculations were conducted using the pure PBEPBE exchange-correlation functional.[32-33] To accurately account for dispersion interactions, the DFT-D3 correction was applied.[34] The calculations were performed with a high plane-wave kinetic energy cutoff of 73 Ry to ensure convergence of the electronic wavefunctions. The Brillouin zone was sampled using a Monkhorst-Pack k-point grid with 27 points. The cell vectors of the periodic systems were fully optimized alongside the atomic coordinates during the geometry reoptimization step to obtain the most stable structures under periodic boundary conditions. All PWDFT calculations were conducted using the Quantum Espresso software package (version 6.0) for the geometry reoptimization of the periodic system geometries, electronic structure research, and the calculation of band gaps.[35-36] The electrical conductivity was derived using the KGEC (Kubo-Greenwood Electrical Conductivity) post-processing code.[37] The method is based on the Kubo-Greenwood formalism, which provides an expression for the frequency-dependent complex conductivity tensor. The general formula for the conductivity σ(ω) is given by:

$$\sigma(\omega) = i \frac{2e^2 \hbar^3}{m_e^2 V} \sum_m \sum_{m'} \frac{\Delta f_{m'm}}{\Delta \epsilon_{mm'}} \frac{\langle m|\nabla|m'\rangle \langle m'|\nabla|m \rangle}{(\Delta \epsilon_{mm'} - \hbar\omega + i\delta/2)},$$

where $m$, $m'$ — single-particle state indices, $\epsilon_m$, $\epsilon_{m'}$ — their energies, $f(\epsilon_m)$, $f(\epsilon_{m'})$ — Fermi–Dirac occupation numbers, $2f(\epsilon m)$ — total KS-orbital occupancy (including spin), $\Delta\epsilon_{mm'} = \epsilon_m - \epsilon_{m'}$, $\Delta f_{mm'} = f(\epsilon_m) - f(\epsilon_{m'})$, $e$ — electron charge, $\hbar$ — Planck's constant, $m_e$ — electron mass, $V$ — system volume, $i\delta/2$ — imaginary damping term. For direct current conductivity (ω→0), the expression simplifies accordingly. The calculation captures intraband, interband, and degenerate-band contributions, and employing a Lorentzian representation of the Dirac delta function.

**Results and Discussion**

Here, we investigate soot samples produced by the electrical breakdown in the PP film capacitors. For modeling, we considered the carbon-to-hydrogen ratio according to the formula $[C_3H_6]_n$ for PP and $[C_{22}H_{10}O_5N_2]_n$ for PI, as well as the established empirical fact that the ratio of electrode thickness to dielectric thickness is on the order of tens of nanometers to micrometers.[38-39]

The calculation of the number of metal and dielectric atoms was performed using the following formula:

$$\frac{n(polymer)}{n(Me)} = \frac{d_{pol\ or\ PI} h_{pol} M_{Me}}{M_{pol} d_{Me} h_{Me}}$$

The terms n(polymer) and n(Me) denote the molar quantities, $d_{pol}$ and $d_{Me}$ the mass densities, $h_{pol}$ and $h_{Me}$ the thicknesses, and $M_{pol}$ and $M_{Me}$ the molar masses of the polymer and metal, respectively. This calculation yielded a molar ratio of dielectric atoms to electrode atoms of approximately 50 to 1. It is important to note that the number of metal atoms was slightly overestimated in this model. This adjustment accounts for the well-known experimental fact that the area of vaporized metal after a breakdown event is typically larger than the affected area of the dielectric because the metals are more destructively affected due to their higher thermal conductivity.[40] In our systems, this metal component was modeled using 3 atoms of zinc and 1 atom of aluminum, simulating approximately a possible Al/Zn alloy. Furthermore, since PP film is often treated (including with oxygen-containing compounds) to enhance its dielectric properties, we conducted our research on the influence of additional oxygen on the properties of the resulting soot. Table 1 presents the studied samples, which contain n(O) = 1, 2, 3, 4, 6, 8, 10. The oxygen-free sample was used as a reference.

Table 1. The parameters of the simulated systems for PP and PI: chemical compositions, the numbers of atomic nuclei, and the numbers of electrons. Note that some elements are simulated with core potentials in PM7.

| # | Composition | # nuclei | # electrons | Abbreviation |
|---|---|---|---|---|
| 1 | $AlZn_3[C_3H_6]_{10}$ | 94 | 343 | 1 Al + 3 Zn + PP |
| 2 | $AlZn_3O[C_3H_6]_{10}$ | 95 | 351 | 1 Al + 3 Zn + PP + 1 [O] |
| 3 | $AlZn_3O_2[C_3H_6]_{10}$ | 96 | 359 | 1 Al + 3 Zn + PP + 2 [O] |
| 4 | $AlZn_3O_3[C_3H_6]_{10}$ | 97 | 367 | 1 Al + 3 Zn + PP + 3 [O] |
| 5 | $AlZn_3O_4[C_3H_6]_{10}$ | 98 | 375 | 1 Al + 3 Zn + PP + 4 [O] |
| 6 | $AlZn_3O_6[C_3H_6]_{10}$ | 100 | 391 | 1 Al + 3 Zn + PP + 6 [O] |
| 7 | $AlZn_3O_8[C_3H_6]_{10}$ | 102 | 407 | 1 Al + 3 Zn + PP + 8 [O] |
| 8 | $AlZn_3O_{10}[C_3H_6]_{10}$ | 104 | 423 | 1 Al + 3 Zn + PP + 10 [O] |
| 9 | $AlZn_3[C_{22}H_{10}O_5N_2]_2$ | 82 | 495 | 1 Al + 3 Zn + PI |
| 10 | $AlZn_3O[C_{22}H_{10}O_5N_2]_2$ | 83 | 503 | 1 Al + 3 Zn + PI + 1 [O] |
| 11 | $AlZn_3O_2[C_{22}H_{10}O_5N_2]_2$ | 84 | 511 | 1 Al + 3 Zn + PI + 2 [O] |
| 12 | $AlZn_3O_3[C_{22}H_{10}O_5N_2]_2$ | 85 | 519 | 1 Al + 3 Zn + PI + 3 [O] |

| 13 | AlZn$_3$O$_4$[C$_{22}$H$_{10}$O$_5$N$_2$]$_2$ | 86 | 527 | 1 Al + 3 Zn + PI + 4 [O] |
| 14 | AlZn$_3$O$_6$[C$_{22}$H$_{10}$O$_5$N$_2$]$_2$ | 88 | 543 | 1 Al + 3 Zn + PI + 6 [O] |
| 15 | AlZn$_3$O$_8$[C$_{22}$H$_{10}$O$_5$N$_2$]$_2$ | 90 | 559 | 1 Al + 3 Zn + PI + 8 [O] |
| 16 | AlZn$_3$O$_{10}$[C$_{22}$H$_{10}$O$_5$N$_2$]$_2$ | 92 | 575 | 1 Al + 3 Zn + PI + 10 [O] |

The initial geometries of all systems were atomized soot samples, into which oxygen atoms were introduced. This structure corresponds to the state of the material after electrical breakdown, when the metallized polymer vaporizes. In real capacitors, the temperature during a breakdown event can reach 5000 to 7000 K. In our model, the initial temperature was set to 5000 K and then decreased to 300 K over 200 steps, with a time step of 0.5 fs. After the system reached 300 K, the kinetic energy injection method was applied. This technique involves introducing portions of kinetic energy into the system to induce structural perturbations. These perturbations enable the system to efficiently overcome potential energy barriers and reach lower-energy states. The method is effective for exploring the potential energy surface of the system and identifying configurations corresponding to global and local minima. Amid successive kinetic energy injections, atomic motion was simulated using PM7-MD. Figure 1 presents the standard enthalpies of formation for the 1Al+3Zn+PP, 1Al+3Zn+PP+1[O], 1Al+3Zn+PI, and 1Al+3Zn+PI+4[O] samples.

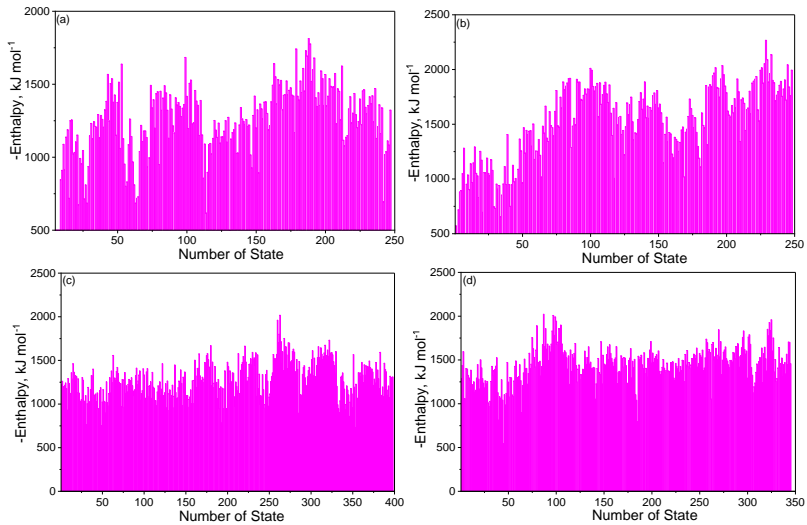

Figure 1. Distribution of the formation energies for the stationary states of (a) 1 Al + 3 Zn + PP + 0 [O], (b) 1 Al + 3 Zn + PP + 1 [O], (c) 1 Al + 3 Zn + PI + 0 [O], and (d)

1 Al + 3 Zn + PI + 4 [O] systems. The enthalpies of formation were inverted for the most stable state to correspond to the highest bar.

Figure 2 shows the supercell, comprising eight unit cells, which was obtained by optimizing the lattice parameters and atomic coordinates for the 1 Al + 3 Zn + PP + 2 [O] system. The structural patterns observed are characteristic of all systems containing PP. Aluminum primarily interacts with hydrogen, forming aluminum hydride $AlH_3$. Zinc atoms incorporate into the carbon chain, favoring bonds with other zinc atoms via hydrogen bridges, forming -Zn-H…Zn- configurations. Similar structural motifs have been observed in our previous studies.[5-6] The gas phase consists predominantly of $H_2$ and CO molecules. Furthermore, the formation of aromatic compounds is also observed.

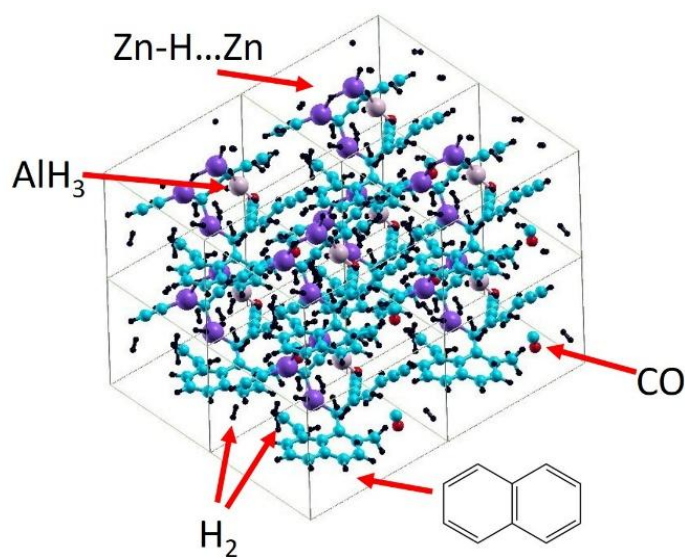

Figure 2. The simulated supercell of the 1 Al + 3 Zn + PP + 2 [O] system. Note that two extra oxygen atoms are covalently bonded to carbon atoms. The geometry of the periodic system was obtained via pure DFT. The carbon atoms are cyan, the oxygen atoms are red, the zinc atoms are violet, the aluminum atoms are pink, and the hydrogen atoms are black.

Figure 3 presents the supercell for the 1 Al + 3 Zn + PI + 6 [O] system. This configuration represents the global minimum structure identified through the simulations. The final geometry, including both the cell shape and atomic positions, was refined via PWDFT. The structural patterns shown are characteristic of most PI-containing structures. In systems with PI, the molar fraction of hydrogen is lower (0.32) compared to PP systems (0.64). Consequently, there is insufficient hydrogen for aluminum to form a hydride. Instead, structures with -C-AlH$_2$ motifs prevail. Zinc atoms incorporate into carbon chains or interact with other zinc atoms directly, forming -C-Zn-Zn-C- configurations without hydrogen bridges. Nitrogen tends to terminate carbon chains, forming -C-N edges. The gas phase is dominated by CO and H$_2$ molecules.

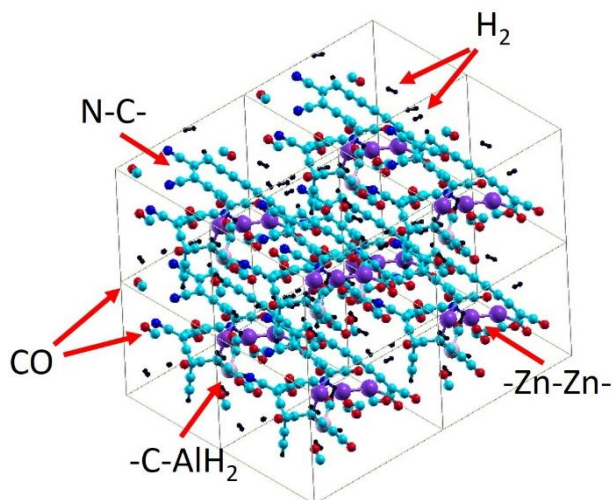

Figure 3. The simulated supercell of the 1 Al + 3 Zn + PI + 6 [O] system. Note that six extra oxygen atoms are covalently bonded to carbon atoms. The geometry was obtained via pure density functional theory. The carbon atoms are cyan, the oxygen atoms are red, the nitrogen atoms are blue, the zinc atoms are violet, the aluminum atoms are pink, and the hydrogen atoms are black.

A critical factor governing the self-healing efficacy is the chemical composition of the decomposition products resulting from the dielectric breakdown. To elucidate the pyrolysis products of PP in the presence of the metal electrode, the global minimum energy structures for each simulated system were analyzed. Table 2 details the composition of the gaseous and solid phases derived from these global minimum configurations for the PP-based systems. The analysis reveals that the gas phase is predominantly composed of H$_2$ molecules, which constitute the largest share and vary in quantity from 19 to 23 molecules across the different systems. Furthermore, CO is present in all systems except the oxygen-free reference, and its yield increases systematically with the initial oxygen content in the system. A maximum of four CO molecules is observed for systems containing 6 [O], 8 [O], and 10 [O]. Unsaturated hydrocarbons, such as C$_2$H$_2$, C$_4$H$_2$, and more exotic species, emerge. $CH_4$ formation is observed in the system containing 8 [O]. In the soot phase, incorporated [O] exhibits a preference for terminating the carbon chains.

Table 2 lists the gas and soot species obtained from the PP systems. Hydrogen molecules are the dominant gas species. CO is generated with the increase of oxygen content, while hydrocarbons gradually decrease. In the case of PP systems without oxygen addition, the products include $H_2$ and $C_2H_2$, but no CO. When 1 and 2 [O] are introduced, CO formation begins, the amount of $H_2$ fluctuates slightly, and $C_2H_2$ oxidizes to CO. No other hydrocarbons are generated, indicating a weak oxidation reaction. When the number of oxygen atoms ranges from 3 to 10, the CO content increases from 1 to 4, and the type or content of hydrocarbons decreases. $H_2$ remains the dominant gaseous product. 19 through 23 $H_2$ molecules with minor fluctuations are observed. This is mainly due to the oxidation of carbon in PP to CO. Meanwhile, hydrocarbons exhibit diversity: $C_2H_2$ appears when oxygen atoms are 3 or 6, $C_4H_2$ at 4 [O], $CH_4$ at 8 [O], and $C_3O_2$ at 10 [O]. Twenty-one molecules of $H_2$ form as the number of oxygen atoms equals ten. $H_2$ is consistently the main gaseous product. Hydrogen atoms of PP are released as $H_2$. Oxidation into water in these chemical compositions is not favored.

Regarding soot products, their elemental composition gradually incorporates more oxygen as its content rises, causing the C/H ratio to adjust with fluctuations. When the oxygen content is zero, the soot consists of $AlZn_3C_{26}H_{18}$. When more than two oxygen atoms are present, the soot begins to incorporate oxygen. Its content generally increases with further oxygen addition, indicating that oxygen participates in the oxidative modification of soot by forming oxygen-containing functional groups. The number of carbon atoms in soot samples changes with increasing oxygen content: it starts at 26 atoms in the absence of oxygen, increases to 29 atoms when 1-2 oxygen atoms are present, then declines to 24-25 atoms, reaching a minimum of 23 carbon atoms when the system contains 10 [O]. Overall, the C/H ratio fluctuates between 1.2 and 2.1, without an obvious monotonic trend. This behavior may be related to the competition between PP cracking and soot dehydrogenation reactions. Metal oxides do not form. Zinc and aluminum are incorporated into the soot structure.

Table 2. Gas and soot species in the PP systems. Three zinc atoms and one aluminum atom represent MFC electrodes. The results are obtained from the global minimum structures of each soot composition.

| Polymer | Gas species | Soot species |
| --- | --- | --- |
| 1 Al+3 Zn + PP | 19 $H_2$, 2 $C_2H_2$ | $AlZn_3C_{26}H_{18}$ |
| 1 Al+3 Zn + PP + 1 [O] | 23 $H_2$, 1 CO | $AlZn_3C_{29}H_{14}$ |
| 1 Al+3 Zn + PP + 2 [O] | 20 $H_2$, 1 CO | $AlZn_3C_{29}H_{20}O$ |
| 1 Al+3 Zn + PP + 3 [O] | 22 $H_2$, 2 CO, 1 $C_2H_2$ | $AlZn_3C_{26}H_{14}O$ |
| 1 Al+3 Zn + PP + 4 [O] | 21 $H_2$, 2 CO, 1 $C_4H_2$ | $AlZn_3C_{24}H_{16}O_2$ |
| 1 Al+3 Zn + PP + 6 [O] | 23 $H_2$, 4 CO, 1 $C_2H_2$ | $AlZn_3C_{24}H_{12}O_2$ |
| 1 Al+3 Zn + PP + 8 [O] | 21 $H_2$, 4 CO, 1 $CH_4$ | $AlZn_3C_{25}H_{14}O_4$ |
| 1 Al+3 Zn + PP + 10 [O] | 21 $H_2$, 4 CO, 1 $C_3O_2$ | $AlZn_3C_{23}H_{18}O_4$ |

The mass fraction of the gas and the soot is shown in Figure 4. When the number of oxygen atoms increases from 1 to 10, the gas mass fraction shows a significant upward trend. With 1–2 [O], the gas mass fraction is around 10 %. When the number of oxygen atoms reaches ten, the gas mass fraction rises to 28 %. This exhibits an approximately linear positive correlation, indicating that the introduction of oxygen atoms promotes the formation of gas species in the system. In contrast to the gas species, the soot mass fraction shows a distinct downward trend with the increase in the number of oxygen atoms: Without oxygen, the soot mass fraction is 86 %; When the oxygen atom count reaches ten, the soot mass fraction decreases to approximately 72 %. It also presents an approximately linear negative correlation, demonstrating that the increase in oxygen atoms inhibits the formation of the soot species. Table 2 reveals that oxygen atoms support oxidation reactions with carbon-based precursors in the system. Carbon atoms of the soot transition into volatile species, carbon-containing gases.

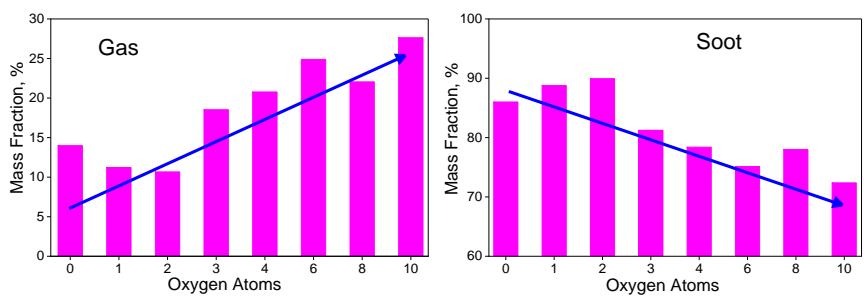

Figure 4. Mass fractions of the gas (left) and the soot (right) versus the number of [O] in the PP systems.

The gas and soot species of PI systems are obtained as listed in Table 3. For the gas species, when no oxygen is present, the products include $H_2$ and CO, but no hydrocarbons or nitrogen-containing gases. This reflects the endogenous oxidation effect of the inherent PI oxygen-containing structure. At one or two [O], $H_2$ forms incrementally without other impurity gases, indicating that the release of $H_2$ out of PI is restricted by its molecular structure. The promoting effect of oxidation on $H_2$ generation is limited. At 3 to 10 [O], CO formation increases, and $H_2$ content fluctuates. $C_2H_2$ was detected at 3 [O]. $H_2O$ and HCN were observed at 4 [O].

The essence of CO dominating the gaseous products lies in the PI molecular structure: PI contains -CO-N- and aromatic rings, with some intrinsic oxygen content. H atoms in PI are bonded to aromatic rings or imide bonds, which have high bond energy. During pyrolysis/oxidation, H atoms are difficult to detach and form $H_2$. The $H_2$ content is generally low and fluctuates, which is significantly different from the PP system. This reflects the decisive influence of polymer molecular structure (aromatic rings vs. alkane chains) on the release pathway of H elements. The appearance of HCN in the presence of four oxygen atoms indicates that imide bonds in PI undergo cracking under medium oxygen conditions. When six or more [O] are present, HCN is absent. The simulation results show no formation of gaseous nitrogen-containing species besides HCN, including $N_2$ and NO.

Al and Zn metal atoms do not participate in oxidation by [O] because they form a thermodynamically stable intermetallic structure. The number of carbon atoms fluctuates and decreases from 37 in the absence of oxygen to 32-33 with 8, 9, and 10 [O]. This result is consistent with the increasing CO generation.

Most nitrogen atoms of PI are retained in the soot. The number of [O] from 3 with no added oxygen to 8 at 10 [O], generally increasing with the addition of exogenous oxygen. This suggests that oxygen participates in the oxidation of the PI soot, forming oxygen-containing functional groups -OH and -C=O. The number of H atoms decreases from 14 at zero [O] to 4 at 6 [O] and rebounds to 10 at 10 [O]. The changes in the soot are in agreement with the total amount of hydrogen molecules released.

Table 3. Gas and soot species in the PI systems. Three zinc and one aluminum atom represent the destroyed electrode. The results have been obtained from the global minimum structures of the PI soot.

| Polymer | Gas species | Soot species |
| --- | --- | --- |
| 1 Al + 3 Zn + PI | 3 $H_2$, 7 CO | $AlZn_3C_{37}N_4H_{14}O_3$ |
| 1 Al + 3 Zn + PI +1 [O] | 6 $H_2$, 5 CO | $AlZn_3C_{39}N_4H_8O_6$ |
| 1 Al + 3 Zn + PI +2 [O] | 7 $H_2$, 9 CO | $AlZn_3C_{35}N_4H_6O_3$ |
| 1 Al + 3 Zn + PI +3 [O] | 4 $H_2$, 8 CO, 1 $C_2H_2$ | $AlZn_3C_{34}N_4H_{10}O_5$ |
| 1 Al + 3 Zn + PI +4 [O] | 5 $H_2$, 11 CO, 1 $H_2O$, 1 HCN | $AlZn_3C_{32}N_3H_7O_2$ |
| 1 Al + 3 Zn + PI +6 [O] | 8 $H_2$, 11 CO | $AlZn_3C_{33}N_4H_4O_5$ |
| 1 Al + 3 Zn + PI +8 [O] | 6 $H_2$, 12 CO | $AlZn_3C_{32}N_4H_8O_6$ |
| 1 Al + 3 Zn + PI +10 [O] | 5 $H_2$, 12 CO | $AlZn_3C_{32}N_4H_{10}O_8$ |

Upon the number of [O] increasing from 0 to 4, the gas mass fraction shows significant fluctuations, with a general tendency to increase. As n(O)=4, the gas mass fraction remains largely stable. Regarding the soot mass fraction, it exhibits an overall fluctuating downward trend. As n(O)=4 exceeds four, the soot fraction stays unchanged. As [O] content grows, it promotes the formation of gas species and suppresses the generation of the carbonaceous soot.

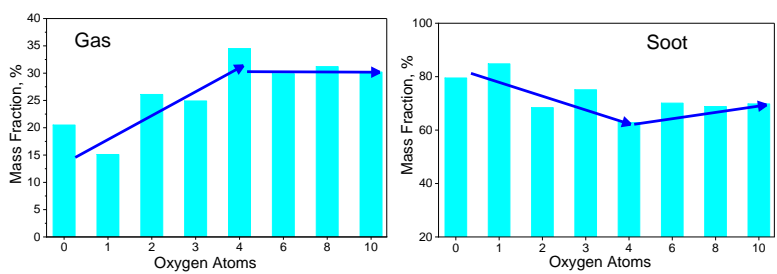

Figure 5. Mass fractions of gas (left) and soot (right) versus the number of oxygen atoms in the PI systems.

The variation in electrical conductivity of the PP soot samples with different oxygen atom contents is shown in Figure 6. It can be seen that the electrical conductivity decreases as the number of oxygen atoms increases. In the absence of oxygen, the electrical conductivity reaches a peak, 2.9 kS/m. As the number of oxygen atoms increases from 0 to 10, the electrical conductivity decreases uniformly, eventually dropping to 1.0 kS/m. The most significant drop in conductivity occurs when the first oxygen atom is introduced, from 2.9 to 2.0 kS/m. After the number of oxygen atoms exceeds six, the rate of conductivity decline slows down (decreasing from 1.8 to 1.0 kS/m), which may be due to the surface oxygen-containing groups approaching saturation, so their effect on the conductivity is minimized.

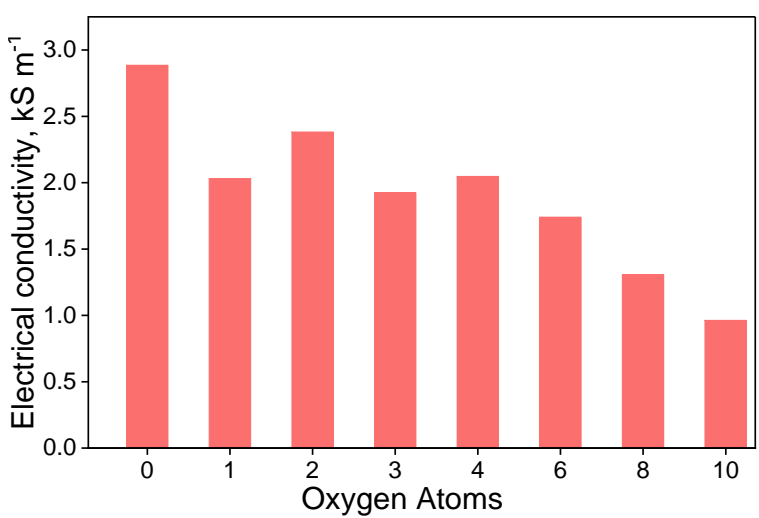

Figure 6. Electrical conductivity of the PP-based soot samples versus the number of [O].

Compared with the electrical conductivity of PP, the variation of PI electrical conductivity with the number of oxygen atoms is shown in Figure 7. As the number of oxygen atoms increases from 0 to 3, the electrical conductivity decreases significantly, dropping to 1.5 kS/m as 3 [O] atoms are involved. This indicates that the introduction of a small number of oxygen atoms will notably weaken the electrical conductivity of the PI soot. As [O] content increases from 3 to 10, electrical conductivity shows some fluctuation around 1.5-1.9 kS/m. In no specimen, conductivity reaches 2.7 kS/m in the non-ozonated PI insulator. This observation suggests that introducing a larger quantity of oxygen atoms cannot reverse the decline in its conductivity. Oxygen atoms alter the molecular structure of the PI soot, like disrupting the conjugated system, which impairs charge transport efficiency.

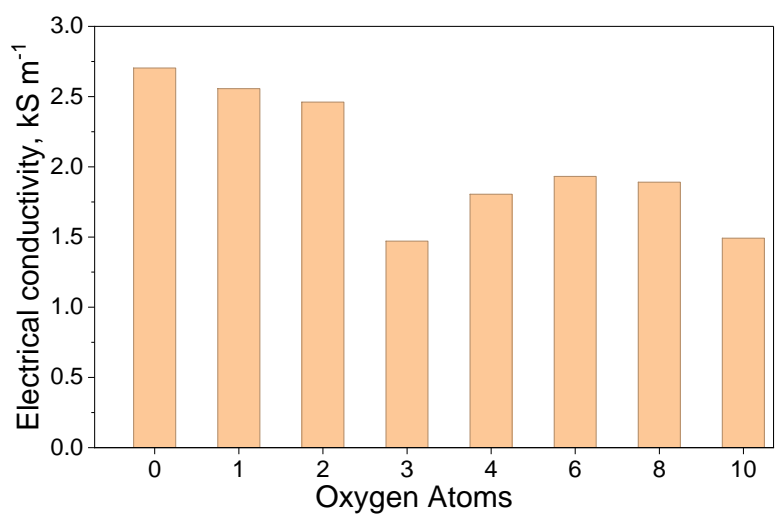

Figure 7. Electrical conductivity in the PI soot samples versus the number of [O].

The lowest unoccupied molecular orbital (LUMO) energy levels, highest occupied molecular orbital (HOMO) energy levels, and band gaps of PP and PI systems with different oxygen atom contents are listed in Tables 5-6. The band gaps in the PI systems are generally narrower than those in the PP systems. The PP systems exhibit a band gap range of 0.411–1.406 eV, while the PI systems have a band gap range of 0.121–0.756 eV.

For the PP systems, the HOMO energy levels show significant fluctuations, 0.136–2.847 eV. The LUMO energy levels are widely scattered, 0.687–4.175 eV, reflecting the high reconstructability of the carbon skeleton after alkane chain decomposition and the significant regulatory effect of oxygen-containing groups on the orbital energy levels. In contrast, the PI systems display lower overall HOMO, 0.003–1.537 eV, and LUMO, 0.274–1.864 eV, energy levels with a narrower fluctuation range. This indicates that the stabilizing effect of the aromatic structure in PI on the frontier orbital energies is not fully destroyed with the introduction of [O]. The LUMO energy in PP + 4 [O] reaches 4.175 eV. The addition of [O] does not induce a monotonic trend in the band gaps of the PP soot samples but generally broadens them. The band gaps of most oxygen-richest PP systems are larger than those of the PP systems, which is 0.637 eV.

Table 5. The LUMO and HOMO energy levels, and the band gaps in the PP systems.

| Composition | LUMO, eV | HOMO, eV | Bandgap, eV |
|---|---|---|---|
| 1 Al + 3 Zn + PP | 3.156 | 2.519 | 0.637 |
| 1 Al + 3 Zn + PP + 1 [O] | 0.884 | 0.151 | 0.733 |
| 1 Al + 3 Zn + PP + 2 [O] | 2.621 | 2.210 | 0.411 |
| 1 Al + 3 Zn + PP + 3 [O] | 2.019 | 0.613 | 1.406 |
| 1 Al + 3 Zn + PP + 4 [O] | 4.175 | 2.847 | 1.328 |
| 1 Al + 3 Zn + PP + 6 [O] | 0.687 | 0.136 | 0.551 |
| 1 Al + 3 Zn + PP + 8 [O] | 2.392 | 1.462 | 0.93 |
| 1 Al + 3 Zn + PP + 10 [O] | 1.414 | 0.136 | 1.278 |

For the PP and PI systems, the variation of the bandgap exhibits a non-monotonic fluctuating characteristic, which reflects that the oxygen content has a certain regulatory effect on the electronic structure. In the PP system, the band gap ranges from 0.411eV to 1.406 eV, with a span of 0.995 eV. In the PI system, the span is 0.635 eV. Owing to their narrower band gaps, the PI systems exhibit superior electrical conductivity. In contrast, the PP systems are more sensitive to [O] content, leading to band gap broadening and electrical conductivity deterioration.

Table 6. The LUMO and HOMO energy levels, and the band gaps in the polyimide contained systems.

| Composition | LUMO, eV | HOMO, eV | Band gap, eV |
|---|---|---|---|

| 1 Al + 3 Zn + PI | 0.939 | 0.789 | 0.150 |
| --- | --- | --- | --- |
| 1 Al + 3 Zn + PI + 1 [O] | 1.291 | 0.916 | 0.375 |
| 1 Al + 3 Zn + PI + 2 [O] | 1.658 | 1.537 | 0.121 |
| 1 Al + 3 Zn + PI + 3 [O] | 0.707 | 0.530 | 0.177 |
| 1 Al + 3 Zn + PI + 4 [O] | 1.864 | 1.439 | 0.425 |
| 1 Al + 3 Zn + PI + 6 [O] | 1.011 | 0.255 | 0.756 |
| 1 Al + 3 Zn + PI + 8 [O] | 0.274 | 0.003 | 0.271 |
| 1 Al + 3 Zn + PI + 10 [O] | 1.413 | 0.962 | 0.451 |

**Conclusions and Final Considerations**

The oxygen content's regulatory role in self-healing of MFCs was investigated in terms of gasified soot fraction, electrical conductivity, and soot band gap. The results show that oxygen oxidizes carbon to CO and suppresses hydrocarbon formation in the PP system. In turn, PI systems use not only [O] from ozonation but also inherent imide bond oxygen for endogenous oxidation. Sporadic emergence of HCN reflects imide bond cleavage. Yet, most nitrogen atoms remain with the carbon atoms in the soot. In the PP systems, [O] from ozonation linearly elevates gas mass fraction from 10 % to 28 % and reduces soot fraction from 86 % to 72 %. The PI soot produces a higher gas fraction at low levels of ozonation, 20 %, but its further ozonation is less effective.

In terms of electrical conductivity, the PP soot conductivity declines uniformly from 2.9 kS/m to 1.0 kS/m. The PI soot conductivity drops to 1.5 kS/m in the 3 [O] system but increases again up to 1.9 kS/m at higher [O] ozonation levels. The PI soot exhibits narrower band gaps (0.121–0.756 eV) than the PP soot (0.411–1.406 eV). This is in line with higher electrical conductivity. Oxygen non-monotonically tailors band gaps in both polymers, while PP is more oxygen-sensitive.

The theoretical, first principles-based work hereby quantified the positive role of insulator ozonation to boost self-healing efficacies in both PP and PI, as exemplified by modest, i.e., realistic, incorporated oxygen contents. The results are addressed to electrical engineers seeking advanced insulator formulations.

## Acknowledgments

The results of the work were obtained using the computational resources of Peter the Great Saint-Petersburg Polytechnic University super-computing center (www.spbstu.ru).

## Conflict of Interest

The authors hereby declare no financial interests and professional connections that might bias the interpretations of the obtained results.

## Credit author statement

N.A.A. Conceptualization; Methodology Development; Validation; Formal analysis; Investigation; Resources; Data Curation; Writing - Original Draft; Writing - Review & Editing; Visualization Preparation; Supervision; Project administration. C.L.: Investigation. V.V.C.: Conceptualization; Methodology Development; Validation; Formal analysis; Investigation; Resources; Data Curation; Writing - Original Draft; Writing - Review & Editing; Visualization Preparation; Supervision; Project administration.

## References

1. Andreeva, N. A.; Chaban, V. V. Self-healing metalized film capacitors: Quo Vadis? Energy Storage and Conversion, **2025**, 3 (2), 2945.

2. Mao, P.; Wang, J.; Zhang, L.; Sun, Q.; Liu, X.; He, L.; Liu, S.; Zhang, S.; Gong, H. Tunable dielectric polarization and breakdown behavior for high energy storage capability in P(VDF-TrFE-CFE)/PVDF polymer blended composite films. Phys Chem Chem Phys, **2020**, 22 (23), 13143-13153, 10.1039/d0cp01071e.

3. Bossa, G. V.; Caetano, D. L. Z. Differential capacitance of curved electrodes: role of hydration interactions and charge regulation. Phys Chem Chem Phys, **2024**, 26 (23), 16774-16781, 10.1039/d4cp00372a.

4. Pan, L.; Wang, F.; He, Y.; Sun, X.; Du, G.; Zhou, Q.; Zhang, J.; Zhang, Z.; Li, J. Reassessing Self-Healing in Metallized Film Capacitors: A Focus on Safety and Damage Analysis. IEEE Transactions on Dielectrics and Electrical Insulation, **2024**, 31 (4), 1666-1675, 10.1109/TDEI.2024.3357441.

5. Chaban, V. V.; Andreeva, N. A. Insulator and electrode materials marginally influence carbonized layer conductivity in metalized-film capacitors. Phys Chem Chem Phys, **2025**, 27 (28), 15154-15162, 10.1039/D5CP00835B.

6. Andreeva, N. A.; Chaban, V. V. Self-healing in dielectric capacitors: a universal method to computationally rate newly introduced energy storage designs. Phys Chem Chem Phys, **2024**, 26 (47), 29393-29405, 10.1039/D4CP03988B.

7. Chaban, V. V.; Andreeva, N. A. Compositional variation in self-healing of Au-electrode dielectric capacitors. Mater Chem Phys, **2025**, 344, 131155, https://doi.org/10.1016/j.matchemphys.2025.131155.

8. He, Q.; Sun, K.; Shi, Z.; Liu, Y.; Fan, R. Polymer dielectrics for capacitive energy storage: From theories, materials to industrial capacitors. Mater Today, **2023**, 68, 298-333, 10.1016/j.mattod.2023.07.023.

9. Gennady E Zaikov, S. K. R. Ozonation of Organic and Polymer Compounds. Rapra Technology Ltd, **2009**, 422 pages. isbn: 1847351433.

10. Xie, J.; Zhao, X.; Zheng, S.; Zhong, S.; Liu, X.; Zhang, M.; Sun, S. All-organic PVDF-based composite films with high energy density and efficiency synergistically tailored by MMA-co-GMA copolymer and cyanoethylated cellulose. Phys Chem Chem Phys, **2023**, 25 (32), 21307-21316, 10.1039/d3cp03007e.

11. Boxue Du , M. X. Polypropylene Film for HVDC Capacitors. Springer Singapore, **2025**, XI, 338 pages. isbn: 978-981-96-3028-8.

12. Du, G.; Zhang, J. Capacitance Evaluation of Metallized Polypropylene Film Capacitors Considering Cumulative Self-Healing Damage. Electronics (Switzerland), **2024**, 13 (14), 2886, 10.3390/electronics13142886.

13. Liu, H.; Du, B. X.; Xiao, M. Improved Energy Density and Charge Discharge Efficiency of Polypropylene Capacitor Film Based on Surface Grafting. IEEE Transactions on Dielectrics and Electrical Insulation, **2021**, 28 (5), 1539-1546, 10.1109/TDEI.2021.009577.

14. Du, B. X.; Xing, J. W.; Ran, Z. Y.; Liu, H. L.; Xiao, M.; Li, J. Dielectric and Energy Storage Properties of Polypropylene by Deashing Method for DC Polymer Film Capacitors. IEEE Transactions on Dielectrics and Electrical Insulation, **2021**, 28 (6), 1917-1924, 10.1109/TDEI.2021.009688.

15. Xie, D.; Min, D.; Huang, Y.; Li, S.; Nazir, M. T.; Phung, B. T. Classified effects of nanofillers on DC breakdown and partial discharge resistance of polypropylene/alumina nanocomposites. IEEE Transactions on Dielectrics and Electrical Insulation, **2019**, 26 (3), 698-705, 10.1109/TDEI.2018.007610.

16. Gao, J.; Ju, H.; Yao, Z.; Zhang, G.; Jiang, Q.; Guo, H. Effect of zinc oxide nanoparticle size on the dielectric properties of polypropylene-based nanocomposites. Polymer Engineering & Science, **2023**, 63 (7), 2204-2217, https://doi.org/10.1002/pen.26370.

17. Xiao, M.; Wang, K.; Song, Y.; Du, B. Dielectric performance improvement of polypropylene film modified by γ-ray irradiation for HVDC capacitors. Journal of Physics D: Applied Physics, **2024**, 57 (12), 125503, 10.1088/1361-6463/ad13ca.

18. Xiao, M.; Zhang, Z.; Du, B.; Wang, B.; Cao, J. Enhanced Dielectric Performance of Polypropylene Film by Surfacing Grafting and Crosslinking for Metalized Film Capacitors. IEEE Transactions on Applied Superconductivity, **2024**, 34 (8), 1-4, 10.1109/TASC.2024.3484329.

19. Ren, W.; Yang, M.; Guo, M.; Zhou, L.; Pan, J.; Xiao, Y.; Xu, E.; Nan, C. W.; Shen, Y. Metallized stacked polymer film capacitors for high-temperature capacitive energy storage. Energy Storage Materials, **2024**, 65, 103095, 10.1016/j.ensm.2023.103095.

20. Huang, X.; Zhang, S.; Zhang, Y.; Zhang, H.; Yang, X. Sulfonated polyimide/chitosan composite membranes for a vanadium redox flow battery: influence of the sulfonation degree of the sulfonated polyimide. Polym J, **2016**, 48 (8), 905-918, 10.1038/pj.2016.42.

21. Mandal, A. K.; Bisoi, S.; Banerjee, S. Effect of Phosphaphenanthrene Skeleton in Sulfonated Polyimides for Proton Exchange Membrane Application. ACS Appl Polymer Mat, **2019**, 1 (4), 893-905, 10.1021/acsapm.9b00128.

22. YIN Yuhang , Z. G., SONG Jingfu, DING Qingjun. High temperature tribological properties of polyimide composites modified by multi-components. Acta Materiae Compositae Sinica, **2022**, 39 (12), 5699-5710.

23. Woo, Y.; Oh, S. Y.; Kang, Y. S.; Jung, B. Synthesis and characterization of sulfonated polyimide membranes for direct methanol fuel cell. J Membr Sci, **2003**, 220 (1), 31-45, https://doi.org/10.1016/S0376-7388(03)00185-6.

24. Andreeva, N. A.; Chaban, V. V. Magnesium perfluorinated pinacolatoborate in diglyme: understanding microscopic structures in rechargeable magnesium batteries. Phys Chem Chem Phys, **2025**, 27 (43), 23116-23126, 10.1039/D5CP02632F.

25. Chaban, V. V.; Andreeva, N. A. Magnesium-based electrolytes with ionic liquids chloride additives: Quantum chemical stationary point analysis. J Mol Liq, **2024**, 402, 124804, https://doi.org/10.1016/j.molliq.2024.124804.

26. Chaban, V. V.; Andreeva, N. A. Mutual miscibility of diethyl sulfoxide and acetonitrile: Fundamental origin. J Mol Liq, **2022**, 349, 118110, 10.1016/j.molliq.2021.118110.

27. Chaban, V. V.; Andreeva, N. A. Mixtures of Diethyl Sulfoxide and Methanol: Structure and Thermodynamics. J Solut Chem, **2022**, 51 (7), 788-801, 10.1007/s10953-022-01167-x.

28. Tortai, J. H.; Bonifaci, N.; Denat, A.; Trassy, C. Diagnostic of the self-healing of metallized polypropylene film by modeling of the broadening emission lines of aluminum emitted by plasma discharge. J Appl Phys, **2005**, 97 (5), 053304, 10.1063/1.1858872.

29. Chaban, V. V.; Andreeva, N. A. Higher hydrogen fractions in dielectric polymers boost self-healing in electrical capacitors. Phys Chem Chem Phys, **2023**, 26 (4), 3184-3196, 10.1039/d3cp05355e.

30. Humphrey, W.; Dalke, A.; Schulten, K. VMD: visual molecular dynamics. J Mol Graph, **1996**, 14 (1), 33-38, 27-38, 10.1016/0263-7855(96)00018-5.

31. Kokalj, A. XCrySDen—a new program for displaying crystalline structures and electron densities. J Mol Graph Model, **1999**, 17 (3), 176-179, https://doi.org/10.1016/S1093-3263(99)00028-5.

32. Yun, Y.; Wu, S.; Wang, D.; Luo, X.; Chen, J.; Wang, G.; Takao, A.; Wan, L. Molecular dynamics simulations in semiconductor material processing: A comprehensive review. Meas J Int Meas Confed, **2025**, 241, 115708, 10.1016/j.measurement.2024.115708.

33. Perdew, J. P.; Burke, K.; Ernzerhof, M. Generalized gradient approximation made simple. Phys Rev Lett, **1996**, 77 (18), 3865-3868, DOI 10.1103/PhysRevLett.77.3865.

34. Grimme, S.; Antony, J.; Ehrlich, S.; Krieg, H. A consistent and accurate ab initio parametrization of density functional dispersion correction (DFT-D) for the 94 elements H-Pu. J Chem Phys, **2010**, 132 (15), 154104, 10.1063/1.3382344.

35. Giannozzi, P.; Baroni, S.; Bonini, N.; Calandra, M.; Car, R.; Cavazzoni, C.; Ceresoli, D.; Chiarotti, G. L.; Cococcioni, M.; Dabo, I.; Dal Corso, A.; de Gironcoli, S.; Fabris, S.; Fratesi, G.; Gebauer, R.; Gerstmann, U.; Gougoussis, C.; Kokalj, A.; Lazzeri, M.; Martin-Samos, L.; Marzari, N.; Mauri, F.; Mazzarello, R.; Paolini, S.; Pasquarello, A.; Paulatto, L.; Sbraccia, C.; Scandolo, S.; Sclauzero, G.; Seitsonen, A. P.; Smogunov, A.; Umari, P.; Wentzcovitch, R. M. QUANTUM ESPRESSO: a modular and open-source software project for quantum simulations of materials. Journal of Physics: Condensed Matter, **2009**, 21 (39), 395502, 10.1088/0953-8984/21/39/395502.

36. Giannozzi, P.; Andreussi, O.; Brumme, T.; Bunau, O.; Buongiorno Nardelli, M.; Calandra, M.; Car, R.; Cavazzoni, C.; Ceresoli, D.; Cococcioni, M.; Colonna, N.; Carnimeo, I.; Dal Corso, A.; de Gironcoli, S.; Delugas, P.; DiStasio, R. A., Jr.; Ferretti, A.; Floris, A.; Fratesi, G.; Fugallo, G.; Gebauer, R.; Gerstmann, U.; Giustino, F.; Gorni, T.; Jia, J.; Kawamura, M.; Ko, H. Y.; Kokalj, A.; Kucukbenli, E.; Lazzeri, M.; Marsili, M.; Marzari, N.; Mauri, F.; Nguyen, N. L.; Nguyen, H. V.; Otero-de-la-Roza, A.; Paulatto, L.; Ponce, S.; Rocca, D.; Sabatini, R.; Santra, B.; Schlipf, M.; Seitsonen, A. P.; Smogunov, A.; Timrov, I.; Thonhauser, T.; Umari, P.; Vast, N.; Wu, X.; Baroni, S. Advanced capabilities for materials modelling with Quantum ESPRESSO. J Phys Condens Matter, **2017**, 29 (46), 465901, 10.1088/1361-648X/aa8f79.

37. Calderín, L.; Karasiev, V. V.; Trickey, S. B. Kubo–Greenwood electrical conductivity formulation and implementation for projector augmented wave datasets. Comput Phys Commun, **2017**, 221, 118-142, https://doi.org/10.1016/j.cpc.2017.08.008.

38. Li, H.; Qiu, T.; Li, Z.; Lin, F.; Wang, Y. ESR Modeling for Atmospheric Corrosion Behavior of Metallized Film Capacitors. IEEE Transactions on Device and Materials Reliability, **2023**, 23 (4), 486-493, 10.1109/TDMR.2023.3309914.

39. Rochefort, C.; Venet, P.; Clerc, G.; Sari, A.; Wang, M. X.; Mitova, R. Thermal Runaway Indicators of Metallized Polypropylene Film Capacitors. IEEE Transactions on Dielectrics and Electrical Insulation, **2024**, 31 (4), 2136-2143, 10.1109/TDEI.2024.3385351.

40. Wu, Z.; Liu, J.; Qi, H.; Liu, S.; Zhong, S. L.; Wang, J.; Dang, Z. M.; Wang, W. Dynamic Process of Self-Healing in Metallized Film: From Experiment to Theoretical Model. IEEE Trans Plasma Sci, **2024**, 52 (3), 780-789, 10.1109/TPS.2024.3366246.